\documentclass[sensors,article,submit,moreauthors,pdftex,10pt,a4paper]{Definitions/mdpi}
\firstpage{1}
\pubvolume{xx}
\issuenum{1}
\articlenumber{1}
\pubyear{2018}
\copyrightyear{2018}

\usepackage[utf8]{inputenc}
\usepackage{mathtools}
\usepackage{algorithmic}
\usepackage[ruled]{algorithm2e}
\usepackage{amsmath,amssymb,amsfonts}
\usepackage{algorithmic}
\usepackage{graphicx}
\usepackage{epstopdf}
\usepackage{color}
\usepackage{subfigure}
\renewcommand{\textcolor}[2]{#2}
\usepackage{textcomp}
\usepackage{array}
\usepackage{float}
\usepackage{url}
\usepackage{hyperref }
\usepackage{xcolor}
\usepackage{colortbl}
\usepackage{epstopdf}
\nolinenumbers

\Title{Optimizing Age of Information in Internet of Vehicles Over Error-Prone Channels%\\
}
\nolinenumbers

\Author{Cui Zhang$^{1}$, Maoxin Ji$^{2}$, Qiong Wu $^{2,*}$ , Pingyi Fan$^{3}$ and Qiang Fan$^{4}$}

\address{
	$^{1}$ \quad School of Internet of Things Engineering, Wuxi Institute of Technology, wuxi, 214121, China; email: faircas85@163.com (C.Z.)\\
	$^{2}$ \quad School of Internet of Things Engineering, Jiangnan University, Wuxi, 214122, China; email: maoxinji@stu.jiangnan.edu.cn (M.J.); qiongwu@jiangnan.edu.cn (Q.W.)\\
	$^{3}$ \quad Department of Electronic Engineering, Beijing National Research Center for Information Science and Technology, Tsinghua University, Beijing 100084, China; email: fpy@tsinghua.edu.cn (P.F.)\\
	$^{4}$ \quad Qualcomm, San Jose, CA 95110, USA;  email: qf9898@gmail.com (Q.F.)
}	
\corres{Correspondence: qiongwu@jiangnan.edu.cn (Q.W.); Tel.: +86-0510-8591-0633 (Q.W.)}
\abstract{
	In the Internet of Vehicles (IoV), Age of Information (AoI) has become a vital performance metric for evaluating the freshness of information in communication systems. Although many studies aim to minimize the average AoI of the system through optimized resource scheduling schemes, they often fail to adequately consider the queue characteristics. Moreover, the vehicle mobility leads to rapid changes in network topology and channel conditions, making it difficult to accurately reflect the unique characteristics of vehicles with  the calculated AoI under ideal channel conditions. 
	This paper examines the impact of Doppler shifts caused by vehicle speeds on data transmission in error-prone channels. Based on the M/M/1 and D/M/1 queuing theory models, we derive expressions for the Age of Information and optimize the system’s average AoI by adjusting the data extraction rates of vehicles (which affect system utilization). We propose an online optimization algorithm that dynamically adjusts the vehicles' data extraction rates based on environmental changes to ensure optimal AoI.
	Simulation results have demonstrated that adjusting the data extraction rates of vehicles can significantly reduce the system’s AoI. Additionally, in the network scenario of this work, the AoI of the D/M/1 system is lower than that of the M/M/1 system.
}
\nolinenumbers
\nolinenumbers

\begin{document}
\nolinenumbers
\section{Introduction}
Currently, vehicular network technology is evolving towards ultra-low latency and ultra-high reliability\cite{997330, 10695132, 10536013}. The diverse array of real-time vehicle applications is underpinned by vehicular network technologies\cite{9597473}. However, in practical applications, vehicular networks still face numerous challenges.

In the context of the Internet of Vehicles (IoV), vehicles often struggle to independently collect sufficient data \cite{10274112, 10615794}, and their computation resources frequently fail to meet task demands\cite{10643168}. Consequently, they rely on the IoV for data processing \cite{clancy2024wireless, moradi2023dsrc, 8672604, 8907851}. However, frequent data exchanges between multiple vehicles and roadside units or base stations (BSs) incur significant resource consumption and even network congestion \cite{10387423, 10032660, 9877926,  9439054}. Optimizing link scheduling can effectively alleviate this issue \cite{5586664, 10163760, 10620366}. Nevertheless, traditional metrics based on communication rate or reliability do not comprehensively reflect the communication performance of the system. Researchers urgently need a new standard to measure the freshness of information, which has led to the emergence of the Age of Information (AoI) concept \cite{kaul2011minimizing}.

AoI refers to the time interval from when specific information is generated in a communication system to when it is actually received and processed by the recipient. This concept is crucial in wireless communication, sensor networks, and real-time data transmission\cite{10663259}. In communication systems, the update frequency of information typically depends on its freshness, which can be measured by AoI. Lower AoI indicates more real-time and fresher information. Therefore, reducing AoI helps improve communication system performance by reducing transmission delays and enhancing real-time capabilities.

Furthermore, the AoI concept plays a vital role in optimizing data transmission strategies, ensuring the timely delivery of critical information. If the AoI exceeds a certain threshold, it may not provide effective information for high real-time requirement tasks, potentially leading to erroneous judgments. Additionally, AoI can help to estimate the information change during the data transmission. For example, in autonomous driving scenarios, vehicle positions change over time, and AoI management aids in more precise vehicle control, ensuring road safety. By managing and minimizing AoI, the system can be improved to meet the demands of real-time communication and data transmission.

In the communication process of IoV, vehicles and base stations exchange large amounts of data\cite{10654286, 10636300}. Base stations provide computation power and service support to vehicles within their coverage\cite{10670125, 10278180}. Resource allocation strategies in the environment significantly affect packet waiting times and transmission intervals\cite{CAO2024103438, 10745538, 10697115, 10699378}. Therefore, designing appropriate resource allocation strategies can help reduce the system's average AoI, which has been extensively studied. However, each base station has a limited service rate; if the request frequency from vehicles is too high, it may increase the number of the requests in the queue, thus impacting the freshness of information received by the base station. Therefore, adjusting the request frequency of vehicles to match the base station's service rate is crucial.

Additionally, recent research often employs ideal channel models, while overlooking real-world channel uncertainties such as signal attenuation and packet collisions. \textcolor{red}{In error-prone channels, when the channel is in an adverse state, packets experience significant interference, leading to transmission failures and necessitating retransmissions. This substantially affects the freshness of information, resulting in higher AoI. Therefore, the consideration of the impact of error-prone channels on AoI helps to better represent the system performance in the AoI model designing. Based on this, this paper introduces an error-prone channel model to accurately reflect actual communication environments, optimize AoI, and enhance the system's real-time response capabilities.}

In this paper, by incorporating an error-prone channel model, we more accurately reflect actual communication environments. We derive an analytical expression for AoI considering both data extraction rate and channel drop probability and prove its convexity, allowing efficient optimization of parameters using convex optimization methods to minimize AoI\footnote{The source code can be found at \url{https://github.com/qiongwu86/Blockchain-Enabled-Variational-Information-Bottleneck-for-Minimizing-AoI-in-IoV}}. The contributions of this paper are summarized as follows:
\begin{itemize}
	\item We model the process of vehicle data extraction and base station service as a Poisson point process, considering an error-prone channel model, and derive an analytical expression for the relationship between vehicle data extraction rate and AoI.
	
	\item Using Python software, we simulate the derived expression(a simple convex function), which can obtain the optimal solution. We analyze the differences in AoI between D/M/1 and M/M/1 systems.
	
	\item Experimental results demonstrate that the proposed method effectively reduces the average age of information and enhances system real-time response capability under various channel conditions.
\end{itemize}

This research fills the gap of ignoring the impact of error-prone channels on AoI, providing new theoretical and practical insights for AoI optimization in complex communication environments.

The rest of this paper is organized as follows: Section II reviews related research work as a foundation for this study. Section III describes the system scenario and channel model, laying the groundwork for subsequent analysis. Section IV derives the analytical expression for AoI under error-prone channel conditions. In Section V, the results are implemented through Python simulations, and the performance of the proposed method is demonstrated through comparative analysis. Section VI summarizes the conclusions of this paper.

\section{Related Work}
Since the AoI was first proposed, it has established a theoretical foundation and has been used as a critical metric in many communication scenarios.

\subsection{Theoretical Research}

\textcolor{red}{The article \cite{9217386} first derived the exact expression for the average Age of Information (AoI) in a single-server multi-source M/M/1 queueing model with a First-Come-First-Served (FCFS) service policy, addressing the inaccuracies in previous calculations of average AoI. The correctness of the theoretical derivation was verified through simulations, but it is limited to the M/M/1 model and may not be applicable in certain scenarios.
The article \cite{10198349} studied the probability and distribution of violations for AoI and PAoI in a multi-source M/G/1/1 preemptive system without buffers. Using time-domain analysis, general formulas for the violation probabilities and probability density functions of these two metrics were derived, and corresponding closed-form expressions were obtained for the M/M/1/1 system. The study showed that as the arrival rate increases, the violation probabilities and variances for each source monotonically decrease. Additionally, the paper proposed an optimization problem for maximizing the violation probability, demonstrating that optimizing the arrival rate distribution can significantly enhance system timeliness.
The article \cite{8820073} investigated the stationary distribution of information age in information update systems, proposing a general formula applicable to a wide range of information update systems. This formula indicates that the stationary distribution of AoI can be represented in terms of system latency and peak AoI's stationary distribution, providing a unified and efficient method for AoI analysis since system latency and peak AoI can be analyzed using standard queuing theory techniques. To demonstrate its applicability, the authors analyzed AoI under four different service policies in a single-server queue: First-Come-First-Served (FCFS), Preemptive Last-Come-First-Served (LCFS), and two non-preemptive LCFS strategies, comparing the average AoI in M/GI/1 and GI/M/1 queues under these service policies.
The article \cite{9347556} analyzed the impact of computation and transmission times on AoI in edge computing environments, proposing a two-phase queue model where the average transmission service time is monotonically dependent on the computation service time. The study covered both non-buffered and single-unit buffered non-preemptive services, as well as preemptive strategies. The authors derived closed-form expressions for average AoI and average peak AoI, revealing the relationships between system parameters and demonstrating the trade-offs between the two through numerical results. The findings indicated that transmission service preemption performs better when the variance of computation time is low, while computation service preemption is advantageous when variance is high. This research significantly extends AoI theoretical analysis beyond earlier studies that were limited to non-preemptive services.}

\textcolor{red}{These works have facilitated the theoretical analysis of AoI; however, they cannot be directly applied to vehicular networks. The high-speed mobility of vehicles leads to rapid changes in channel conditions, and retransmissions caused by this instability can significantly impact the system's AoI. Therefore, it is essential to conduct a deeper analysis of AoI considering the specific characteristics of vehicular networks.}

\subsection{Practical Applications}

\textcolor{red}{The article \cite{10419638} discusses certain technical shortcomings in Cellular Vehicle-to-Everything (C-V2X) technology that may lead to increased information age and affect communication reliability. This paper proposes an enhanced C-V2X Mode 4 and introduces a more suitable performance metric for vehicular networks called decision-age AoI. By optimizing resource scheduling schemes in the C-V2X protocol, the waiting time of packets in the queue can be effectively reduced, optimizing information freshness.
The article \cite{peng2020age} employs a semi-persistent scheduling-based Medium Access Control (MAC) protocol for sidelink communication in vehicular networks. They use AoI to assess the performance of the MAC layer in C-V2X sidelink communication, introducing a collaborative method based on bearing to avoid packet collisions among vehicles, demonstrating significant improvements in traditional metrics like transmission reliability.
The article \cite{9442821} investigates a general AoI penalty function, which can describe any exponential and logarithmic shapes of AoI penalties by adjusting the values of $\alpha$ and $\beta$, thus addressing the complexity of deploying multiple AoI penalty functions and corresponding algorithms in multitasking systems. The correctness of the proposed method is analyzed based on wireless-powered communication networks, using a one-dimensional search algorithm to find the optimal battery capacity, thereby promoting research on AoI applications.
The article \cite{10683037} discusses AoI issues in low Earth orbit networks based on Non-Orthogonal Multiple Access (NOMA), utilizing deep reinforcement learning along with low-complexity power allocation algorithms. An energy-aware AoI algorithm is proposed to optimize AoI performance verified through simulation comparisons.
The article \cite{10193386} considers intelligent transportation systems, optimizing AoI to ensure reliable data transmission between vehicles and roadside units. The paper combines packet delay and throughput to compute the weighted sum of AoI for effective minimization, finding the optimal solution through machine learning’s Alpha-Beta pruning algorithm.
The article \cite{parvini2023aoi} investigates AoI perception in queue systems when addressing wireless resource management problems in vehicular networks. They utilize a distributed resource allocation framework based on multi-agent reinforcement learning, where each queue manager acts as an agent interacting with the environment to learn its optimal strategy. By adjusting resource allocation in the environment, the system's AoI can be minimized.
The article \cite{emara2020mec} finds that the concept of AoI can provide deeper assessments when addressing vulnerable road users in vehicular networks, particularly in safety-critical applications. Based on this, the paper proposes a cellular network architecture that includes Mobile Edge Computing (MEC) infrastructure and compares it with traditional network architectures. Simulation results demonstrate the significant improvement in information age performance of the proposed architecture.
The article \cite{mlika2022deep} employs NOMA to provide minimal AoI and high reliability for vehicular safety information. They reformulate a resource allocation problem based on AoI, including half-duplex transceiver selection, broadcast coverage optimization, power allocation, and resource block scheduling. By utilizing matching algorithms in conjunction with deep deterministic policy methods, they achieve collaborative resource allocation that significantly reduces the system's average AoI.}

\textcolor{red}{These works have deeply integrated the optimization of AoI with specific application environments; however, most of them are based on optimizing scheduling schemes at the MAC layer or reducing packet transmission times through resource allocation adjustments. They have not considered the queue characteristics within the system, including client arrival rates and base station service rates. Furthermore, they overlook the additional delays caused by potential packet transmission failures in error-prone channels, motivating our current work. Building on the above analysis, we have thoroughly combined queuing theory's queue models with the practical conditions in vehicular networks, introducing an error-prone channel model influenced by vehicle speed. We derived a formula for information age and proposed an online optimization algorithm that optimizes the system’s average information age by adjusting the data extraction rates of vehicles.}

\begin{figure}[t]
	\centering
	\includegraphics[width=\columnwidth, trim=0.5cm 0.5cm 0.5cm 0.5cm, clip]{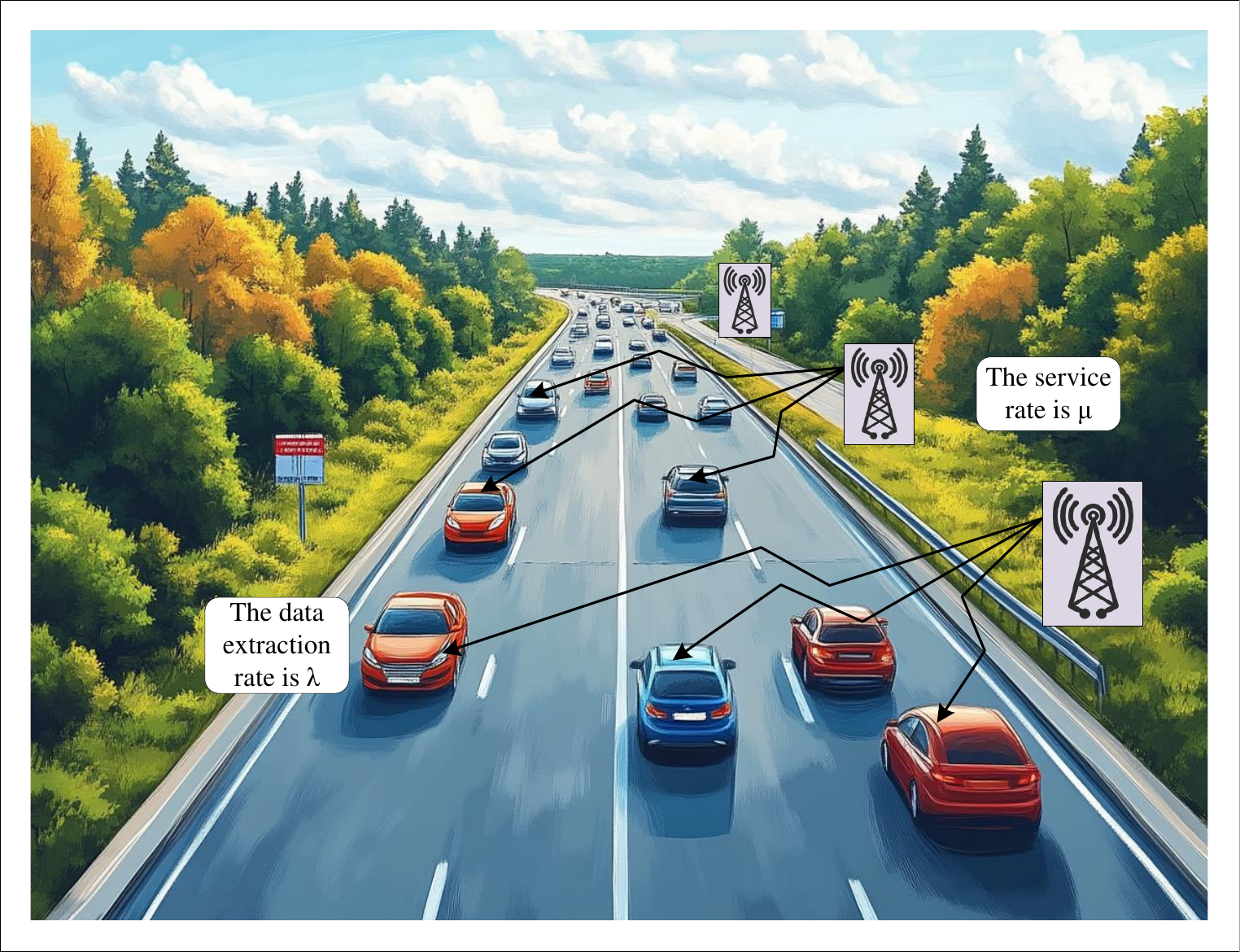}
	\caption{Communication Scenario between Vehicle and Roadside Unit or Base Station in Internet of Vehicles Environment.}
	\label{fig1}
\end{figure}

\section{System Model}

In this section, we construct a vehicular network environment and the data extraction process for vehicles. This will lay the foundation for the subsequent formula derivation.

Fig.~\ref{fig1} illustrates the vehicular network environment we consider. The environment includes various types of vehicles with limited computing capabilities and BS with abundant computing resources, such as roadside units and servers. \textcolor{red}{The vehicles are uniformly distributed on the road, with an initial speed within a certain range}, and base stations are deployed at intervals along the sides of the road. Suppose there are a total of \(N\) base stations in the environment, and each base station can connect to a maximum of \(M\) vehicles. Vehicles communicate with base stations via V2I communication to obtain entertainment information and can complete complex tasks with the assistance of the base stations. \textcolor{red}{It is assumed that the data extraction rate for vehicles is $\lambda$ and the service rate of the base stations is \( \mu \). Here, the data extraction rate of vehicles can be seen as the frequency of vehicles requesting services in the system, while the service rate of the base stations represents the capacity of the base stations to handle tasks. These two parameters are key indicators that affect system performance in a queuing theory model.}

\textcolor{red}{In queuing theory, the arrival of requests is typically modeled as a Poisson process. This means that the number of data requests (such as data extractions) occurring within a fixed time interval follows a Poisson distribution, which is commonly used to model random, independent events occurring at a constant average rate.} Due to the data interaction between base stations and vehicles, the process of extracting data from each vehicle can be modeled as a Poisson point process \(A\) on \( \mathbb{R} \). It has the following two characteristics:
\begin{itemize}
	\item \textcolor{red}{\textbf{Number of events within a finite interval:} For any finite interval \([a, b)\), the number of events (such as data extractions) occurring within this interval can be described by a Poisson random variable. The expected value of this random variable is $\lambda([a, b))$, which acts as a measure for the process $A$ (referred to as a non-negative Radon measure). This measure helps quantify the number of arrivals within the interval \([a, b)\).}
	\item \textcolor{red}{\textbf{Independence between intervals:} If we divide the timeline into several non-overlapping intervals \([a_1,b_1), [a_2,b_2), \ldots, [a_K,b_K)\), the number of events occurring in each interval is independent of the others. The expected number of events in each interval is given by \(\lambda([a_1,b_1)), \lambda([a_2,b_2)), \ldots, \lambda([a_k,b_k))\).}
\end{itemize}

Here, the expression \(\lambda([a, b))\) represents the total number of arrivals within the interval \([a, b)\). Its value is influenced by both the intensity function and the length of the interval: i.e.,  a higher intensity function and a longer interval result in a larger value of \(\lambda\).

For a point process defined on non-negative real numbers $\{A_{(i)}\}_{i\geq 1}$, the points are arranged in ascending order according to their arrival times, i.e., $A_1 \leq A_2 \leq A_3 \leq \ldots$. It is straightforward to see that the inter-arrival times between consecutive points are random variables, denoted as $\mathcal{T}_i = A_i - A_{i-1}$, where $\mathcal{T}_1 = A_1$ and $i=2,3,4,\ldots$. In our system, there exists a concept known as inter-arrival times or arrival delays. Now, we have the following two cases.

CASE \uppercase\expandafter{\romannumeral1} (Homogeneous Poisson Process): \textcolor{red}{$\{A_{(i)}\}_{i\geq 1}$ is a homogeneous Poisson process with a constant rate (also known as the intensity function) $\Lambda$. It is known that the mean of the inter-arrival times $\mathcal{T}_i$ is $1/\Lambda$, and these times are independently and identically distributed (iid) exponential random variables. Therefore, the probability that an event arrives before time t can be expressed as:}
\begin{equation}
	\mathbb{P}(\mathcal{T}_i \leq t) = 1 - e^{-\Lambda t}.
	\label{eqa1}
\end{equation}

CASE \uppercase\expandafter{\romannumeral2} (Non-homogeneous Poisson Process): \textcolor{red}{In a non-homogeneous Poisson process, the intensity function $\Lambda(t)$ is a variable that depends on time.} For $\{A_{(i)}\}_{i \geq 1}$ that is a non-homogeneous Poisson process with intensity $\Lambda(t)$, the corresponding arrival rate $\mathcal{T}_i$ can be estimated by the following method. The distribution of the first arrival time $\mathcal{T}_1 = A_1$ is given by
\begin{equation}
	\mathbb{P}(\mathcal{T}_1 \leq t_1) = 1 - e^{-\int_0^{t_1}e^{-\Lambda (x)dx}}, \label{eqa2}
\end{equation}
then, given the first arrival delay $\mathcal{T}_1 = A_1$, the second conditional arrival time $\mathcal{T}_2$ is,
\begin{equation}
	\mathbb{P}(\mathcal{T}_2 \leq t_2|\mathcal{T}_1 \leq t_1) = 1 - e^{-\int_{t_1}^{t_2}e^{-\Lambda (x)dx}}, \label{eqa3}
\end{equation}
similar, for $i=3,4,...$. In a non-homogeneous Poisson process $\{A_{(i)}\}_{i \geq 1}$ with intensity $\Lambda(t)$, each point is independently distributed in the interval $a \in [0, t)$. \textcolor{red}{Therefore, the probability that an event arrives before time \(a\) can be expressed as follows:}
\begin{equation}
	\mathbb{P}(A_i \leq a) = \frac{\lambda(a)}{\lambda(t)},
	\label{eqa4}
\end{equation}
\textcolor{red}{where,}
\begin{equation}
	\lambda(t) = \lambda([0,t)) = \int_0^t\Lambda(t)dt.
	\label{eqa5}
\end{equation}

Similarly, the service process of the base station follows a Poisson distribution with a rate of \( \mu \).
Table \ref{tab:1} summarizes the parameters presented in the text.

\begin{table*}[ht]
	\centering
	\caption{Summary of Parameters in This Paper}
	\begin{tabular}{|c|c|}
		\hline
		Parameter Name & Symbol \\
		\hline
		Number of base stations & $N$ \\
		\hline
		Maximum number of vehicles per base station & $M$ \\
		\hline
		Vehicle data extraction rate & $\lambda$ \\
		\hline
		Base station service rate & $\mu$ \\
		\hline
		System utilization & $\rho$ \\
		\hline
		Vehicle speed & $v$ \\
		\hline
		Carrier frequency & $f_c$ \\
		\hline
		Doppler shift & $f_d$ \\
		\hline
		Probability of different channel & $P_p$ \\
		\hline
		Probability of ideal channel & $P_i$ \\
		\hline
		Channel drop probability & $p_d$ \\
		\hline
		Collision probability & $p_c$ \\
		\hline
		Average age of information & $\Delta$ \\
		\hline
		\textcolor{red}{Fading margin} & \textcolor{red}{$F$} \\
		\hline
		\textcolor{red}{Queue waiting time} & \textcolor{red}{$W$} \\
		\hline
		\textcolor{red}{Server processing time} & \textcolor{red}{$S$} \\
		\hline
		\textcolor{red}{Total packet time in system} & \textcolor{red}{$T$} \\
		\hline
		\textcolor{red}{D/M/1 arrival time} & \textcolor{red}{$D$} \\
		\hline
		\textcolor{red}{Intensity function} & \textcolor{red}{$\Lambda$} \\
		\hline
		\textcolor{red}{Collision interval} & \textcolor{red}{$\tau_c$} \\
		\hline
	\end{tabular}
	\label{tab:1}
\end{table*}

\section{Age of Information Optimization over Error-Prone Channels} 
In this section, we will discuss the impact of error-prone channels on AoI calculation and derive an expression for AoI. It is cover the impact of vehicle data extraction rates on AoI in the environment through a simple convex optimization method.
\subsection{Channel Model}
In our system, we consider the impact of non-ideal, error-prone channels on information transmission. The age of information helps the base stations or vehicles trace back to the time when the information was generated, and thus reconstructing real-time data through estimation. The key of this approach is to accurately analyze the dynamic characteristics of the underlying cellular network channels, such as the channel dynamics in Vehicle-to-Vehicle (V2V) and V2X LTE. In the subsequent sections, we will provide an approximate analysis method, drawing on reference \cite{TMC_Pokhrel}, which considers the dynamic changes in channel states and fundamental details.

\begin{figure}[t]	
	\centering
	\includegraphics[width=\columnwidth, trim=0.5cm 0.5cm 0.5cm 0.5cm, clip]{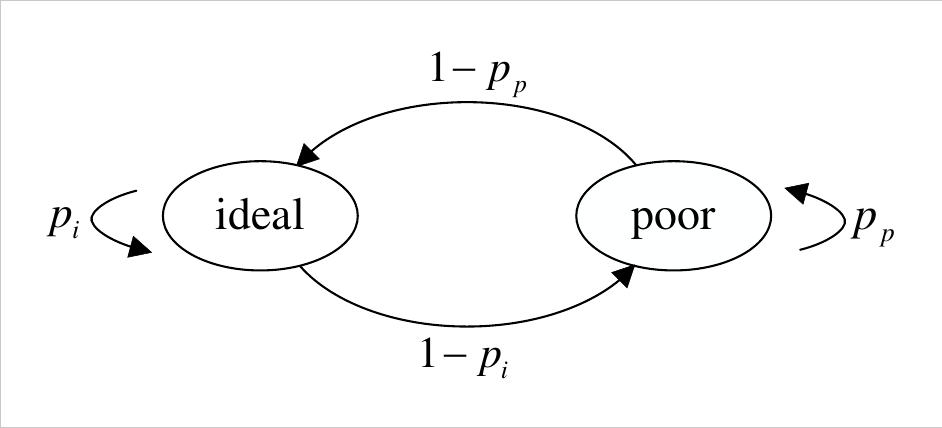}
	\caption{Approximate Cellular Channel State Transition Diagram}
	\label{fig:2}
\end{figure}

We consider discrete time slots where each slot is equal to the  transmission time of a block or module within our system. As depicted in Fig.~\ref{fig:2}, the underlying channel dynamics are assumed to alternate between optimal conditions (where all transmissions succeed) and adverse conditions (where all transmissions fail). \textcolor{red}{In each time slot, if the channel is currently in an optimal state, there is a probability \(p_i\) that it will remain in this state, and a probability \(1-p_i\) that it will switch to an adverse state in the next time slot. Conversely, if the channel is currently in an adverse state, there is a probability \(p_p\) that it will remain in this adverse state in the next time slot, and a probability \(1-p_p\) that it will transition to an optimal state.} In this context, each transmission attempt results in one of two outcomes, either success or failure. Consequently, the likelihood of successful transmission remains constant for each attempt and can be described by a geometric distribution.

Let \(\theta\) denote the maximum rate at which link-layer frames can be transmitted to vehicles, defined as \(\theta = \frac{\text{bit rate}}{\text{frame size in bits}}\). Let \(v\) represent the speed of the vehicle, and \(f_c\) be the carrier frequency. The Doppler frequency is calculated as \(f_d = \frac{f_c v}{c}\), where \(c = 3 \times 10^8 \, \text{m/s}\). 

The channel is considered to be in a poor condition when the received signal-to-noise ratio (SNR) drops below the threshold \(\frac{E[\text{SNR}]}{F}\), where \(F\) is the fading margin, assuming a specific modulation and coding scheme. Conversely, if the SNR remains above this threshold, the channel is regarded as being in an ideal state. 

The average probability of transmission failure due to channel errors is expressed as \(\bar{p}_e = 1 - e^{-1/F}\). Furthermore, the correlation coefficient between two samples of the fading channel amplitude at the Doppler frequency \(f_d\) is given by \(\rho = J_0\left(\frac{2 \pi f_d}{\theta}\right)\), leading to \(\eta = \sqrt{\frac{2}{F(1 - \rho^2)}}\). Here, \(1/\theta\) indicates the time required for frame transmission over the channel, while \(J_0(.)\) represents the zero-order Bessel function. Lastly, the system's static transition probability is defined using a Markov function as described below:
\begin{equation}
	\left\{
	\begin{aligned}
		&P_p = 1 - \frac{\mathbb{Q}(\eta,\rho\eta)-\mathbb{Q}(\eta,\rho\eta)}{e^{1/F}-1}, \\
		&P_i = 1 - \frac{1-\bar{p}_e(2-P_p)}{1-\bar{p}_e}.\\		
	\end{aligned}
	\right.
	\label{pppi}
\end{equation}
The probabilities $P_p$ and $P_i$ represent the probabilities of persisting in poor and ideal channels, respectively. 

In \cite{TCOM_Pokhrel}, the probability of a single update being discarded in the channel is as follows:
\begin{equation}
	p_d=v_L\frac{1-p_i}{2-p_p-p_i}+\frac{l_L}{1+\frac{1-p_i}{1-p_p}},
	\label{pd}
\end{equation}
where $v_L$ is the probability of one frame transmission failure in the poor channel, and $l_L$ is the probability of one frame transmission failure in the ideal channel.

\subsection{AoI computation}
In terms of vehicle data extraction rate and base station service rate, this article first considers the M/M/1 system. In the M/M/1 system, both arrival and service are random and satisfy the property of memorylessness, meaning that future arrival and service times are independent of the past. This implies that each arrival and each service event are mutually independent. Additionally, the complexity and size of vehicle data may affect the efficiency of service process. Specifically, larger data volume may require a longer service time, directly impacting the overall performance of the system. The M/M/1 system is commonly used to study performance metrics such as queue length, customer waiting time, and system busy rate. By analyzing the relationship between arrival and service processes and vehicle data, it is possible to better optimize system performance, reduce waiting times, and improve service quality.

\textcolor{red}{The D/M/1 system is another important model in queuing theory, where “D” indicates that the arrival process is deterministic, while “M” signifies that the service time follows an exponential distribution, still maintaining the memoryless property. “1” indicates that there is only one server in the system. The main feature of the D/M/1 system is that the arrival rate is fixed (for example, a customer arrives at fixed time intervals), while the service times are random and typically can be represented by an exponential distribution. In this model, while the arrival process is deterministic, the randomness of the service process still allows for queuing phenomena. The D/M/1 system can be used to study specific scenarios, such as tasks with scheduled arrival times and their processing under random service times.}

Using queuing theory methods, the following metrics can be achieved such as average queue length, average waiting time, average residence time, and system utilization rate. The analysis of the M/M/1 system is based on the steady-state assumption, meaning that the system is in a stable state and there is no accumulation between arrival and service processes. This model can be applied to various practical scenarios such as call centers, network data transmission, computer servers, etc. 

\textcolor{red}{Furthermore, the D/M/1 system can also compute these key metrics, though its characteristics require corresponding adjustments. In the D/M/1 system, while the arrival process is deterministic, the service time remains random; thus, queuing theory analysis can be used to calculate similar performance metrics. However, the relationship between the arrival rate and the service process will be different. For example, the D/M/1 system may lead to potentially higher queue lengths and waiting times due to the fixed-interval arrival pattern, especially under longer service times.}
	
\textcolor{red}{It is important to note that both the M/M/1 and D/M/1 systems are simplified theoretical models with certain conditions, such as infinite buffer size and distributions of arrival and service rates, which may not fully reflect all scenarios in the real world. Nevertheless, they provide a fundamental framework for understanding and analyzing the performance characteristics of queuing systems.}

For the M/M/1 system, where the vehicle data extraction rate is $\lambda$ and the base station service rate is $\mu$, we have
\begin{equation}
	E\left[ X \right]=1/\lambda.
	\label{tex}
\end{equation}

Also,
\begin{equation}
	E\left[ {{X}^{2}} \right]\text{ }=\text{ 2}/{{\lambda }^{2}}.
	\label{tm1}
\end{equation}

Additionally, the service time is also independently and identically exponentially distributed. The average service time is given by:
\begin{equation}
	E[{{T}_{ser}}]=1/\mu.
	\label{pc1}
\end{equation}

\textcolor{red}{The system utilization $\rho$ can be calculated as follows:}
\begin{equation} 
	\rho =\lambda /\mu.
	\label{pc2}
\end{equation}
Referring to the work of Kual et al.,
\begin{equation}
	{{T}_{i}}={{W}_{i}}+{{S}_{i}},
	\label{tm2}
\end{equation}
\textcolor{red}{where, ${T}_{i}$ represents the total time a packet spends in the system, including both waiting time and service time. ${W}_{i}$ indicates the average waiting time of the packet in the system queue, while $S_i$ denotes the average time required by the server to process the packet.}

However, Kual et al. did not consider the impact of error-prone channels on the system's information age. Therefore, this paper takes the error prone phenomenon into account and gets some new insights. 

In fact, the expected waiting time can be expressed as,
\begin{equation}
	\begin{split}
		E[{{W}_{i}}|{{X}_{i}}=x]
		&=E[{{({{T}_{i-1}}-x)}^{+}}|{{X}_{i}}=x] \\
		& =E[{{(T-x)}^{+}}] \\
		& =\int_{x}^{\infty }{(t-x){{f}_{T}}}(t)dt \\
		& =\frac{{{e}^{-\mu (1-\rho )x}}}{\mu (1-\rho )}.
	\end{split}
\end{equation}
Here, 
\begin{equation}
	{{f}_{T}}(t)=\mu (1-\rho ){{e}^{-\mu (1-\rho )t}},t\ge 0.
	\label{ft}
\end{equation}
\textcolor{red}{where $x$ represents the time point of data arrival.}

As one BS is connected to M vehicles, there is a probability of collision among the data transmissions from these M vehicles. According to the LTE CAT M1 specification, collisions should be avoided by ensuring that consecutive transmissions are 2 to 3 time slots apart. We adopt a spacing of 3 time slots, $\tau_c = 3\tau_t$, where $\tau_t$ is the time slot interval. Therefore, the probability of collision is given by:

\begin{equation}
	\begin{aligned}
		&p_c = 1-\prod^{M} Pr((T_{m_1}-T_{m_2})>\tau_c), \\
		&\forall m_1,m_2 \in [1,M] \quad\&\quad m_1 \neq m_2,
		\label{pc1}
	\end{aligned}	
\end{equation}

where $T_{m_1}$ and $T_{m_2}$ represent the data sending time of two vehicles \textcolor{red}{and $Pr((T_{m_1}-T_{m_2})>\tau_c)$ denotes the probability that m1 and m2 do not collide}. Also, due to \eqref{tm1}, $T_{m_1}$ and $T_{m_2}$ are Poisson process delays plus a constant. Therefore, we have,
\begin{equation} 
	p_c = 1-e^{-\lambda M(M-1) \tau_c/2},
	\label{pc2}
\end{equation}

Considering $p_d$ and $p_c$, \eqref{ft} is updated into,

\begin{equation}
	E[{{W}_{i}}|{{X}_{i}}=x]=\frac{{{e}^{-\mu (1-\rho )x}}}{\mu (1-\rho )}\frac{1}{1-{{p}_{c}}}\frac{1}{1-{{p}_{d}}}.
\end{equation}

\textcolor{red}{By substituting \(p_c\), we obtain:}

\begin{equation}
	E[{{W}_{i}}|{{X}_{i}}=x]=\frac{{{e}^{-\mu (1-\rho )x}}}{\mu (1-\rho )}\frac{{{e}^{\lambda M(M-1){{\tau }_{c}}/2}}}{1-{{p}_{d}}}.
\end{equation}

We can see that the latter term is just a coefficient that is independent of the variable, so we have:

\begin{equation}
	\begin{aligned}
		E[W_i X_i] &= \int_{0}^{\infty} xE[W_i | X_i = x] f_{X_i} dx \\
		& = \frac{\rho}{\mu^2 (1-\rho)} \frac{e^{\lambda M(M-1) \tau_c / 2}}{1-p_d}.  
	\end{aligned}
\end{equation} 

Therefore, equation is is transformed to:
\begin{equation}
	\Delta = \frac{1}{\mu} \left(1 + \frac{1}{\rho} + \frac{\rho^2}{1-\rho} \frac{e^{\lambda M(M-1) \tau_c / 2}}{1-p_d}\right).
	\label{mm1}
\end{equation}

The extremum in \eqref{mm1} cannot be easily obtained through mathematical derivation due to its complexity. However, we can get the extremum by simulation, and its value is around 0.53.

Similarly, in the D/M/1 system, the average information age is calculated as:

\begin{equation}
	\Delta = \frac{1}{\mu} \left[ \frac{1}{2\rho} + \frac{1}{1-\beta} \right].
\end{equation}
Meanwhile, as \( L_X \) is the Laplace transform of the mutual inter-arrival time distribution, we have

\begin{equation}
	\beta = L_X(\mu(1-\beta)),
\end{equation}

The Lambert \( W \) function can be used for calculation,

\begin{equation}
	\beta = e^{-\mu(1-\beta)D} = -\rho W(-\rho^{-1}e^{-1/\rho}).
\end{equation}
\textcolor{red}{where $D$ represents the deterministic arrival time in D/M/1 system.}

Taking collision probability and discard probability into account, equation \ref{mm1} is updated to,
\begin{equation}
	\Delta = \frac{1}{\mu} \left[ 1 + \frac{1}{2\rho} + \frac{\beta}{1-\beta} \frac{e^{DM(M-1)\tau_c/2}}{1-p_d} \right].
\end{equation}

Now we have the analytical expressions for calculating the information age in both M/M/1 and D/M/1 systems in error-prone channels, which can be analyzed through simulations based on the derived equations. \textcolor{red}{Considering the changing number of vehicles in the environment and the service status of the base stations, based on this expression, we can design a real-time online algorithm to optimize the AoI in the communication environment. Since the number of vehicles and service status of base stations are uncontrollable, we can adjust the data extraction rate of vehicles to optimize the AoI. The process of determining the vehicle extraction rate in real-time can be represented as Algorithm \ref{alg:rda}. If vehicles are functioning normally, the optimal data extraction rate is calculated and applied using the formula. When there is a change in the number of vehicles in the environment or in the number of vehicles serviced by each base station, the data extraction rate is recalculated.}

\begin{algorithm}[t]
	\SetKwInOut{Input}{Input}\SetKwInOut{Output}{Output}
	\BlankLine
	\While{vehicles work}{
		\If{$\rho \neq \rho^*$}{
			$\rho  \leftarrow \rho ^*$ by calculating\;
		}
		
		\If{$N \rightarrow N^{new}$}{
			Jump step2\;
		}
		\If{$M \rightarrow M^{new}$}{
			Jump step2\;
		}
	}
	\caption{Real-time data extra and BS server algorithm}
	\label{alg:rda}
\end{algorithm}

\section{Performance Evaluation}
In this section, we use Python as the simulation tool. Based on the expressions derived in the previous section, \textcolor{red}{we analyze the impact of factors such as the number of vehicles in the environment, the channel drop probability, the ratio of vehicle data extraction rate to base station service rate ($\rho$), and collision slots on the AoI. We compare the proposed optimization algorithm with a strategy of randomly selecting $\rho$. In the random strategy, 100 values of $\rho$ are randomly selected, and the system’s average AoI is calculated and averaged.}

\begin{figure}[t]	
	\centering
	\includegraphics[width=3.5in]{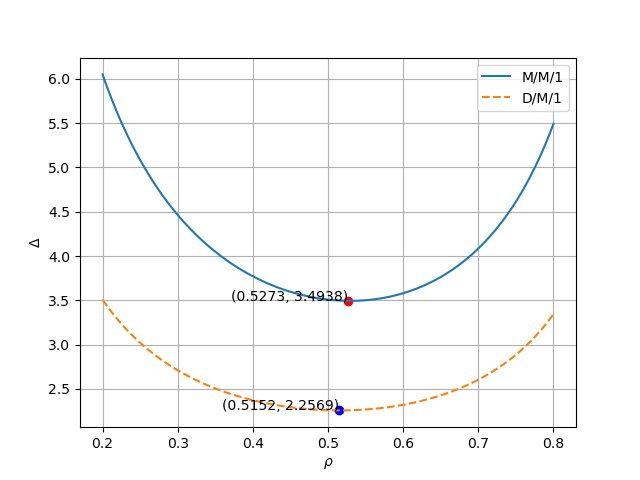}
	\caption{AoI with $\rho$ between M/M/1 and D/M/1.}
	\label{fig:3}
\end{figure}

As shown in Fig.~\ref{fig:3}, \textcolor{red}{we compare the average AoI of D/M/1 and M/M/1 systems with respect to different system utilization rates $\rho$. In practice, since the system service rate $\mu$ is fixed, adjusting the $\rho$ value will tune the vehicle data extraction rate $\lambda$.} It can be seen that the AoI differs between M/M/1 and D/M/1 systems. In the M/M/1 system, the optimal value of \( \rho \) is approximately 0.53, \textcolor{red}{mapping to an average AoI of about 3.49.} 0.2 and 0.53, a larger \( \rho \) results in a smaller age of information.When $\rho$ is 0.53 and 0.8, a larger value leads to a larger age of information. In the D/M/1 system, the optimal value of \( \rho \) is approximately 0.515, \textcolor{red}{corresponding to an average AoI of about 2.26.} when \( \rho \) is between 0.2 and 0.515, it larger value results in a smaller age of information. Between 0.515 and 0.8, a larger \( \rho \) leads to a larger age of information. This is determined by the formula of  the age of information. For the M/M/1 system, Equation (4.17) determines its characteristics, while for the D/M/1 system, Equation (4.24) determines its characteristics.

\begin{figure}[t]	
	\centering
	\includegraphics[width=3.5in]{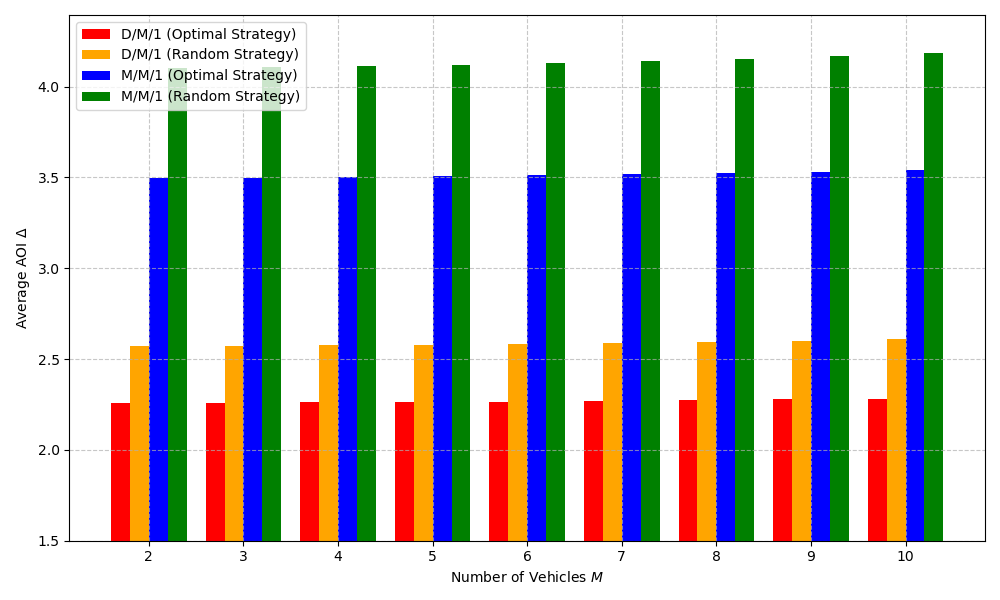}
	\caption{Comparison of Random and Optimal Strategies Under Different Numbers of Vehicles.}
	\label{fig:4}
\end{figure}

\textcolor{red}{Fig.~\ref{fig:4} shows the average AoI caused by using different strategies to determine $\rho$ values in D/M/1 and M/M/1 systems. In the D/M/1 system, the random strategy results in an average age of information that is approximately $15\%$ higher than that of the optimal strategy, whereas in the M/M/1 system, the random strategy results in an average age of information that is approximately $32\%$ higher than that of the optimal strategy. As can be seen, the optimal strategy significantly outperforms the random strategy within the same system. This is because selecting an appropriately optimal data extraction rate that adapts to environmental changes is more effective than randomly selecting a data extraction rate. This confirms the effectiveness of our proposed online optimization method for vehicle data extraction rates.}

\begin{figure}[t]	
	\centering
	\includegraphics[width=3.5in]{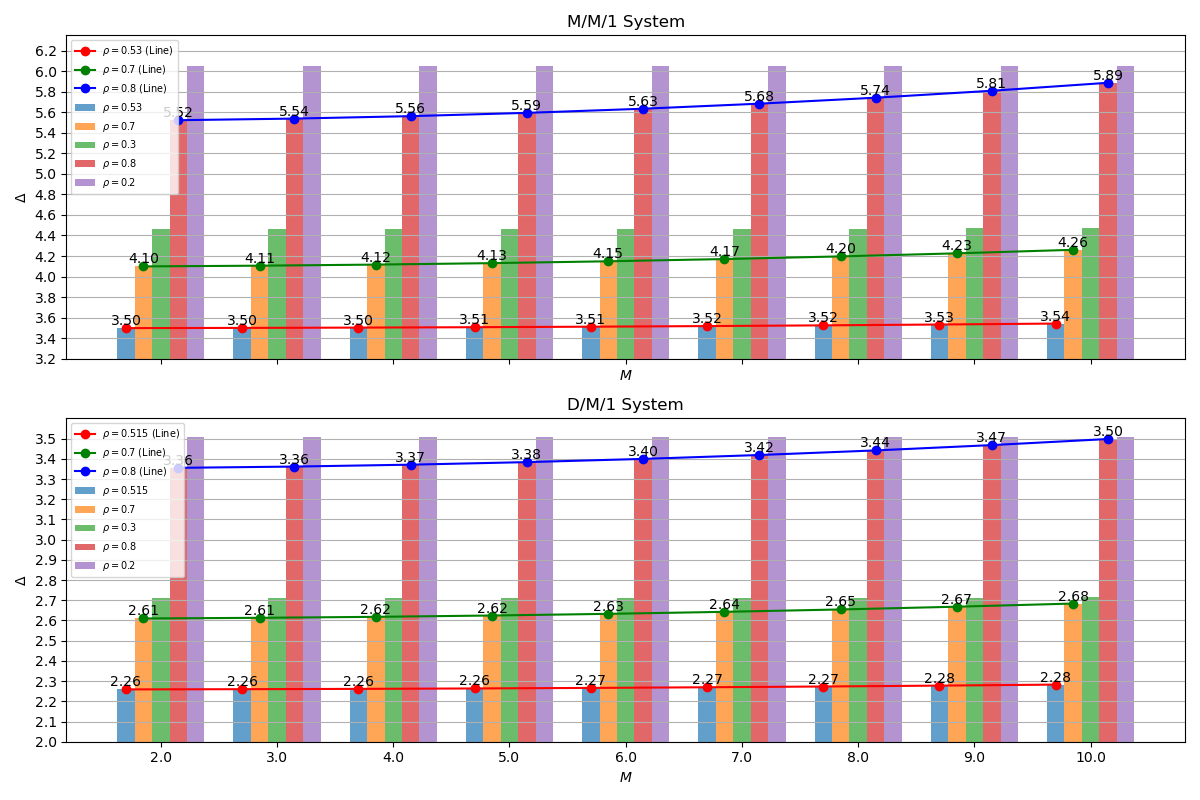}
	\caption{AoI with $M$ between M/M/1 and D/M/1.}
	\label{fig:5}
\end{figure}
Fig.~\ref{fig:5} can be observed that as the maximum number of connected vehicles increases, \textcolor{red}{the AoI shows an increasing trend in both M/M/1 and D/M/1 systems, and this trend becomes more significant as the $\rho$ value increases. In the M/M/1 system, as the number of vehicles increases from 2 to 10, the average AoI for the curve with $\rho = 0.53$ rises by 0.04, for the curve with $\rho = 0.70$ it rises by 0.16, and for the curve with $\rho = 0.80$ it rises by 0.37. In the D/M/1 system, as the number of vehicles increases from 2 to 10, the average AoI for the curve with $\rho = 0.515$ rises by 0.02, for the curve with $\rho = 0.70$ it rises by 0.7, and for the curve with $\rho = 0.80$ it rises by 0.14.} This increase can be attributed to the change in collision probability. \textcolor{red}{The larger the $\rho$ value, the higher the data arrival frequency in the system, making vehicle requests more frequent and more prone to collisions. This impact is greater when the number of vehicles increases.} The increase in collision probability leads to an increase in the time that information waits for transmission in the system, and consequently, the age of information also increases.
This relationship can be further explained as a progressive process. Firstly, increasing the maximum number of connected vehicles leads to an increase in collision probability, which results in an increase in the time that information waits for transmission in the system. As the waiting time increases, the age of information also increases. Therefore, the ultimate result is a compromised performance in terms of information age.

It is worth noting that the D/M/1 system performs better relative to the M/M/1 system. Regardless of how the maximum number of connected vehicles changes on the axes, the information age of the D/M/1 system is smaller than that of the M/M/1 system. This indicates that the D/M/1 system is superior to the M/M/1 system. This may be because the D/M/1 system can handle collisions more effectively and has a lower collision probability, thereby reducing the waiting time for information in the system and ultimately reducing the age of information. 

\begin{figure}[t]	
	\centering
	\includegraphics[width=3.5in]{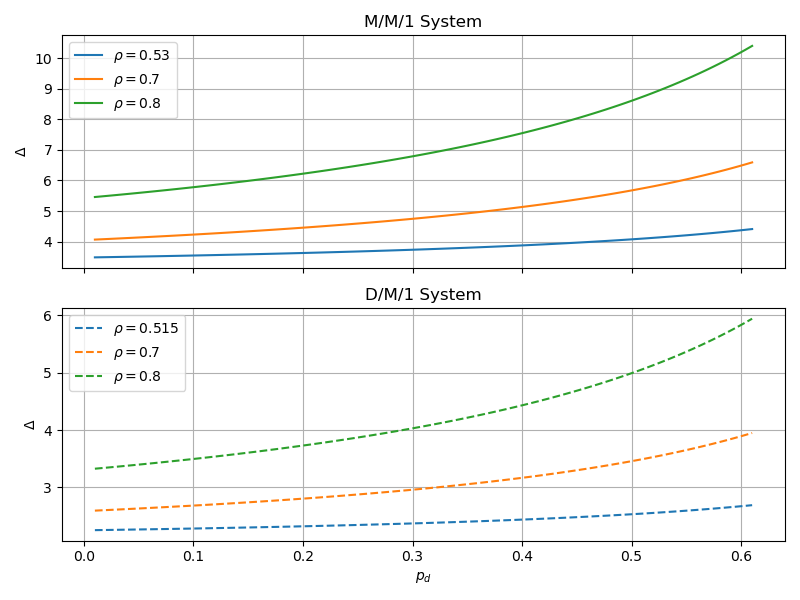}
	\caption{AoI with $P_d$ between M/M/1 and D/M/1.}
	\label{fig:6}
\end{figure}
\textcolor{red}{In Fig.~\ref{fig:6}, the observed curves clearly show that as dropout probability increases, the AoI also shows an increasing trend. This phenomenon is common to both M/M/1 and D/M/1 systems. This is because the dropout probability acts as an indicator reflecting the channel state, and its increasing value will degrade channel congestion, consequently raising the probability of errors.}

An increase in dropout probability leads to an increased number of retransmissions, since information needs to be resent whenever a loss occurs. This extends the time information waits for transmission within the system, thus increasing the AoI. Therefore, the increase in dropout probability directly results in an increase in the AoI.

Further comparing the M/M/1 and D/M/1 systems, it is evident that no matter how the dropout probability changes on the axes, the AoI of the M/M/1 system is consistently greater than that of the D/M/1 system. This indicates that the M/M/1 system is inferior to the D/M/1 system. This phenomenon may be attributed to differences in system complexity. In the M/M/1 system, the data retrieval process is not deterministic but rather more complex, leading to greater instability in the M/M/1 system’s performance as compared to the D/M/1 system. As shown in the figure, with the increase in collision slots, both M/M/1 and D/M/1 systems display a downward trend in AoI. This is because increased collision slots reduce the probability of collisions at adjacent time points when different vehicles send data. \textcolor{red}{Additionally, with the increase in $\rho$, the trend of AoI growth becomes more significant. One possible reason is that the higher data arrival frequency increases collision probability, and another could be that the base station faces greater pressure while providing services, increasing the waiting time for packets to be accepted and consequently increasing the AoI. Therefore, at higher $\rho$ values, the increase in dropout probability may be more sensitive.}

\begin{figure}[t]	
	\centering
	\includegraphics[width=3.5in]{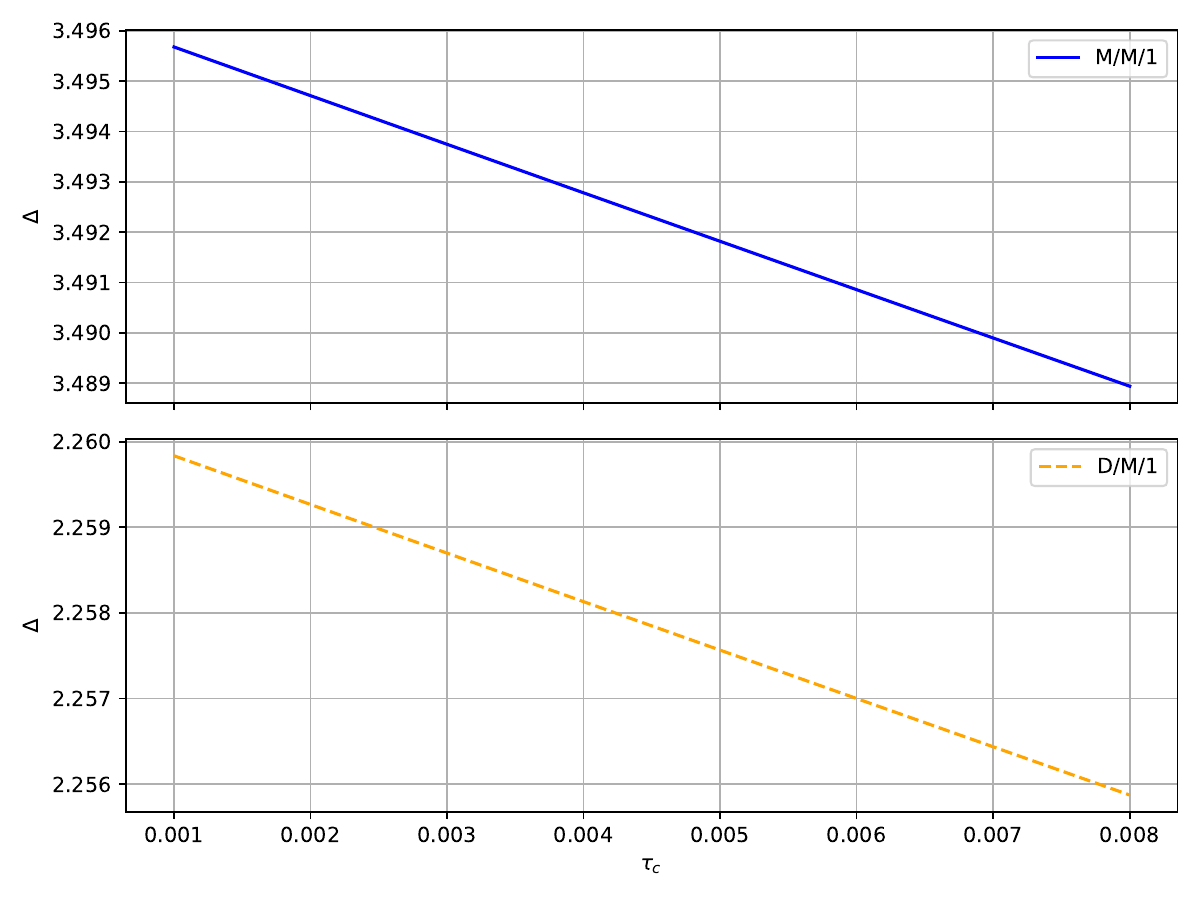}
	\caption{AoI with $\tau_c$ between M/M/1 and D/M/1.}
	\label{fig:7}
\end{figure}

Referring to Fig.~\ref{fig:7}, when collision slots increase, the time intervals between data transmissions among vehicles increase, thus reducing the likelihood of collisions between adjacent vehicles. This lowers the collision probability of the system and reduces the number of information retransmissions. As the number of retransmissions decreases, the total transmission time in the system also decreases. Consequently, the age of information in the system decreases correspondingly.

In summary, the increase of collision slots enables the system to avoid collisions effectively, reducing the number of information retransmissions and thereby decreasing the waiting time in the system, consequently lowering the age of information. This indicates that to some extent, increasing collision slots can improve system performance and reduce information transmission delays. Therefore, the D/M/1 system is more efficient in handling losses and retransmissions, thus exhibiting a lower age of information.

\begin{figure}[t]	
	\centering
	\includegraphics[width=4.5in]{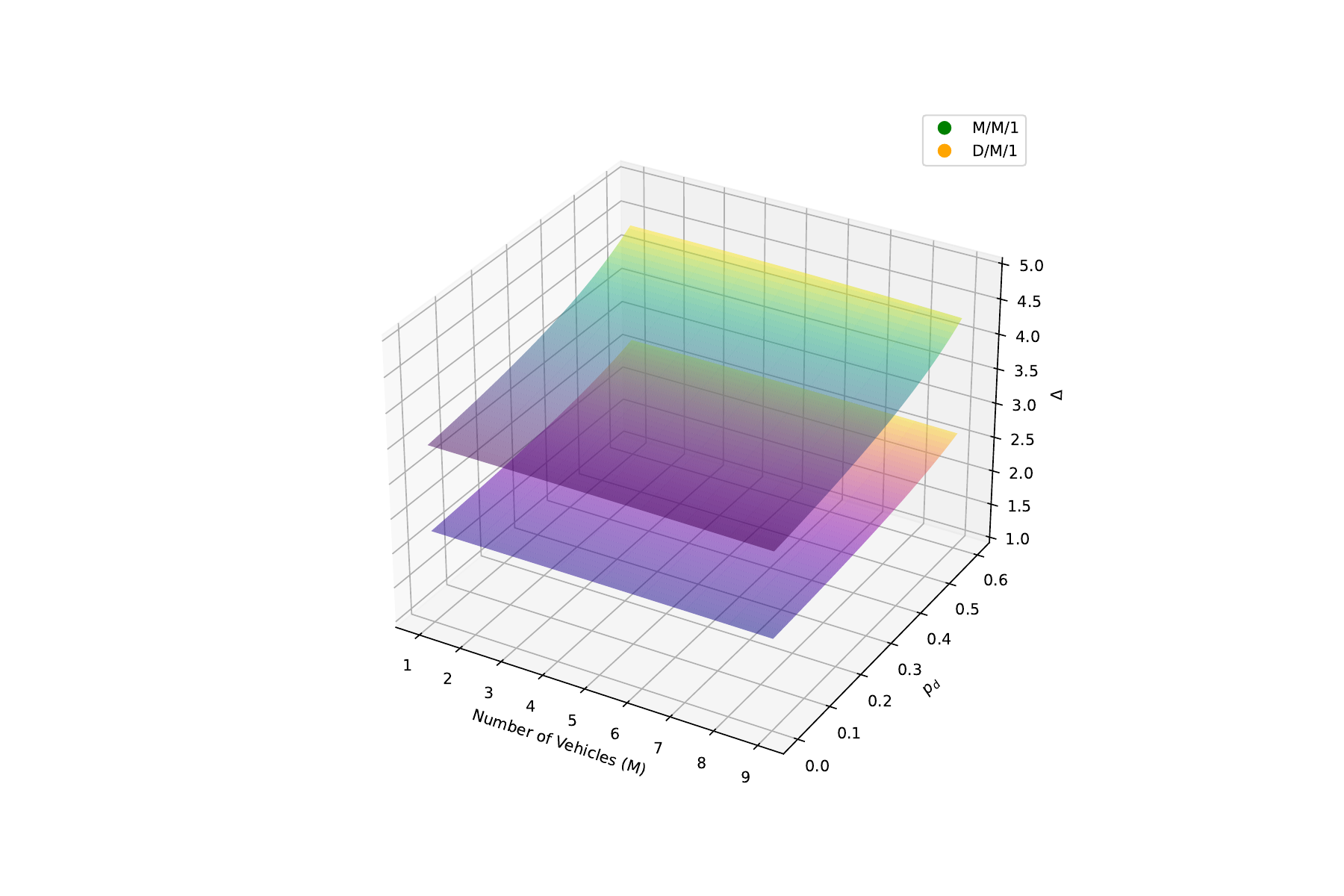}
	\caption{AoI with $m$ and $p_d$ between M/M/1 and D/M/1.}
	\label{fig:8}
\end{figure}

\textcolor{red}{As shown in Fig.~\ref{fig:8}, the AoI increases with an increasing number of vehicles, irrespective of whether it is an M/M/1 or a D/M/1 system. This is because as the number of vehicles increases while the number of base stations remains unchanged, congestion occurs in the queues, which prevents packets from being serviced promptly, and increases AoI. Similarly, the channel dropout probability also affects the AoI, whether in the M/M/1 system or the D/M/1 system. Due to the same reasons, deteriorating channel conditions lead to frequent retransmissions, preventing timely service for packets, thus leading to continuous growth in AoI. This illustrates the impact of vehicle count and base station connectivity on AoI, suggesting the need for effective scheduling and resource allocation strategies when designing vehicular network systems to reduce AoI and improve system performance and efficiency.}

Therefore, in practical applications, optimization and design of vehicular network systems need to comprehensively consider factors such as vehicle count, base station connection count, and queue management to minimize the growth of the age of information and enhance the efficiency and reliability of data transmission. These research findings provide important references and guidance for optimizing vehicular network systems, contributing to the overall performance improvement and user experience enhancement of the systems.

\section{Conclusion}

\textcolor{red}{This paper accurately reflects the actual communication environment in the IoV by considering the error-prone channel conditions caused by Doppler shifts. Based on this, we derived an AoI calculation formula under error-prone channel conditions and optimized the system's AoI by adjusting the data extraction rates of the vehicles. From the simulation results, we have drawn the following conclusions:}

\begin{itemize}
	\item \textcolor{red}{\textbf{Relationship between channel conditions and AoI:} When the channel conditions deteriorate, the channel drop probability increases, which leads to a rise in the system's average AoI. As the data extraction rate of vehicles increases, system utilization also increases, and the impact of channel drop probability becomes more pronounced, resulting in a greater increase in average AoI.}
	
	\item \textcolor{red}{\textbf{Impact of data extraction rates:} The data extraction rate of vehicles significantly affects the system’s average AoI. Both low and high data extraction rates can lead to higher AoI. A low data extraction rate results in low system utilization, causing resource wastage. Conversely, a high data extraction rate increases the system's sensitivity, meaning that an increase in the number of vehicles or a deterioration in channel conditions will lead to a substantial rise in the system's average AoI, and this increase is more evident as system utilization increases.}
	
	\item \textcolor{red}{\textbf{Method validation and system comparison:} Through simulations, we have validated the effectiveness of the proposed method in reducing average AoI and enhancing the system's real-time response capability under various channel conditions. We also analyzed the differences between D/M/1 and M/M/1 systems, and demonstrated that in the studied environment, the D/M/1 system achieves a lower AoI.}
\end{itemize}

\textcolor{red}{These conclusions provide theoretical support and practical guidance for optimizing information freshness in IoV under complex communication environments. However, our study does not delve deeply into the physical layer and largely remains theoretical. In future research, it would be beneficial to consider specific case studies to analyze and optimize the system’s Age of Information. For certain time-varying information, we could explore the use of AoI prediction to obtain real-time information. These aspects are worth further investigation.}

\authorcontributions {Conceptualization, C.Z. and M.J.; Methodology, C.Z., M.J., and Q.W.; Software, C.Z. and M.J.; Writing—Original Draft Preparation, C.Z. and M.J.; Writing—Review and Editing, P.F. and Q.F. All authors have read and agreed to the published version of the manuscript.}

%%%%%%%%%%%%%%%%%%%%%%%%%%%%%%%%%%%%%%%%%%
\funding{This work was supported in part by the National Natural Science Foundation of China under Grant No. 61701197, in part by the National Key Research and Development Program of China under Grant No. 2021YFA1000500(4), in part by the 111 project under Grant No.~B23008.}

\vspace{5pt}
\newcommand{\dataavailability}{%
\noindent\textbf{Data Availability Statement:}
Data are contained within the article.
}
\dataavailability %MDPI: We encourage all authors of articles published in MDPI journals to share their research data. In this section, please provide details regarding where data supporting reported results can be found, including links to publicly archived datasets analyzed or generated during the study. Where no new data were created, or where data is unavailable due to privacy or ethical re-strictions, a statement is still required. Suggested Data Availability Statements are available in section “MDPI Research Data Policies” at \url{https://www.mdpi.com/ethics}.} 
%Author: Exclude this declaration	

\conflictsofinterest{Author Qiang Fan was employed by the company Qualcomm. The remaining authors declare that the research was conducted in the absence of any commercial or financial relationships that could be construed as a potential conflict of interest.}

\reftitle{References}

\vspace{-0.25cm}
\scriptsize

\bibliographystyle{IEEEtran}
\bibliography{tsp}

\begin{thebibliography}{-------}
\providecommand{\natexlab}[1]{#1}

\bibitem[Fan \em{et~al.}(2002)Fan, Feng, Wang, and Ge]{997330}
Fan, P.; Feng, C.; Wang, Y.; Ge, N.
\newblock Investigation of the time-offset-based QoS support with optical burst
  switching in WDM networks.
\newblock  2002 IEEE International Conference on Communications. Conference
  Proceedings. ICC 2002 (Cat. No.02CH37333),  2002, Vol.~5, pp. 2682--2686
  vol.5.
\newblock
  doi:{\changeurlcolor{black}\href{https://doi.org/10.1109/ICC.2002.997330}{\detokenize{10.1109/ICC.2002.997330}}}.

\bibitem[Wang \em{et~al.}(2024)Wang, Wu, Fan, Fan, Zhu, and Wang]{10695132}
Wang, X.; Wu, Q.; Fan, P.; Fan, Q.; Zhu, H.; Wang, J.
\newblock Vehicle Selection for C-V2X Mode 4-Based Federated Edge Learning
  Systems.
\newblock {\em IEEE Systems Journal} {\bf 2024}, pp. 1--12.
\newblock
  doi:{\changeurlcolor{black}\href{https://doi.org/10.1109/JSYST.2024.3459926}{\detokenize{10.1109/JSYST.2024.3459926}}}.

\bibitem[Wu \em{et~al.}(2024)Wu, Wang, Fan, Fan, Zhu, and Letaief]{10536013}
Wu, Q.; Wang, W.; Fan, P.; Fan, Q.; Zhu, H.; Letaief, K.B.
\newblock Cooperative Edge Caching Based on Elastic Federated and Multi-Agent
  Deep Reinforcement Learning in Next-Generation Networks.
\newblock {\em IEEE Transactions on Network and Service Management} {\bf 2024},
  {\em 21},~4179--4196.
\newblock
  doi:{\changeurlcolor{black}\href{https://doi.org/10.1109/TNSM.2024.3403842}{\detokenize{10.1109/TNSM.2024.3403842}}}.

\bibitem[Cao \em{et~al.}(2022)Cao, Zeng, Wang, Sharma, Ma, Liu, and
  Zhou]{9597473}
Cao, D.; Zeng, K.; Wang, J.; Sharma, P.K.; Ma, X.; Liu, Y.; Zhou, S.
\newblock BERT-Based Deep Spatial-Temporal Network for Taxi Demand Prediction.
\newblock {\em IEEE Transactions on Intelligent Transportation Systems} {\bf
  2022}, {\em 23},~9442--9454.
\newblock
  doi:{\changeurlcolor{black}\href{https://doi.org/10.1109/TITS.2021.3122114}{\detokenize{10.1109/TITS.2021.3122114}}}.

\bibitem[Luo \em{et~al.}(2024)Luo, Shao, Cheng, Zhou, Zhang, Yuan, and
  Li]{10274112}
Luo, G.; Shao, C.; Cheng, N.; Zhou, H.; Zhang, H.; Yuan, Q.; Li, J.
\newblock EdgeCooper: Network-Aware Cooperative LiDAR Perception for Enhanced
  Vehicular Awareness.
\newblock {\em IEEE Journal on Selected Areas in Communications} {\bf 2024},
  {\em 42},~207--222.
\newblock
  doi:{\changeurlcolor{black}\href{https://doi.org/10.1109/JSAC.2023.3322764}{\detokenize{10.1109/JSAC.2023.3322764}}}.

\bibitem[Zhao \em{et~al.}(2024)Zhao, Hou, Huang, Lin, and Shan]{10615794}
Zhao, Y.; Hou, F.; Huang, J.; Lin, B.; Shan, H.
\newblock Delay Optimization in Vehicular Edge Computing with Sensing
  Information Fusion and Heterogeneous Tasks.
\newblock  2024 IEEE International Conference on Communications Workshops (ICC
  Workshops),  2024, pp. 1888--1894.
\newblock
  doi:{\changeurlcolor{black}\href{https://doi.org/10.1109/ICCWorkshops59551.2024.10615794}{\detokenize{10.1109/ICCWorkshops59551.2024.10615794}}}.

\bibitem[Zhang \em{et~al.}(2024)Zhang, Zhang, Wu, Fan, Fan, Wang, and
  Letaief]{10643168}
Zhang, C.; Zhang, W.; Wu, Q.; Fan, P.; Fan, Q.; Wang, J.; Letaief, K.B.
\newblock Distributed Deep Reinforcement Learning Based Gradient Quantization
  for Federated Learning Enabled Vehicle Edge Computing.
\newblock {\em IEEE Internet of Things Journal} {\bf 2024}, pp. 1--1.
\newblock
  doi:{\changeurlcolor{black}\href{https://doi.org/10.1109/JIOT.2024.3447036}{\detokenize{10.1109/JIOT.2024.3447036}}}.

\bibitem[Clancy \em{et~al.}(2024)Clancy, Mullins, Deegan, Horgan, Ward, Eising,
  Denny, Jones, and Glavin]{clancy2024wireless}
Clancy, J.; Mullins, D.; Deegan, B.; Horgan, J.; Ward, E.; Eising, C.; Denny,
  P.; Jones, E.; Glavin, M.
\newblock Wireless Access for V2X Communications: Research, Challenges and
  Opportunities.
\newblock {\em IEEE Communications Surveys \& Tutorials} {\bf 2024}, {\em
  26},~2082--2119.
\newblock
  doi:{\changeurlcolor{black}\href{https://doi.org/10.1109/COMST.2024.3384132}{\detokenize{10.1109/COMST.2024.3384132}}}.

\bibitem[Moradi-Pari \em{et~al.}(2023)Moradi-Pari, Tian, Bahramgiri, Rajab, and
  Bai]{moradi2023dsrc}
Moradi-Pari, E.; Tian, D.; Bahramgiri, M.; Rajab, S.; Bai, S.
\newblock DSRC versus LTE-V2X: Empirical performance analysis of direct
  vehicular communication technologies.
\newblock {\em IEEE Transactions on Intelligent Transportation Systems} {\bf
  2023}, {\em 24},~4889--4903.

\bibitem[Cheng \em{et~al.}(2019)Cheng, Lyu, Quan, Zhou, He, Shi, and
  Shen]{8672604}
Cheng, N.; Lyu, F.; Quan, W.; Zhou, C.; He, H.; Shi, W.; Shen, X.
\newblock Space/Aerial-Assisted Computing Offloading for IoT Applications: A
  Learning-Based Approach.
\newblock {\em IEEE Journal on Selected Areas in Communications} {\bf 2019},
  {\em 37},~1117--1129.
\newblock
  doi:{\changeurlcolor{black}\href{https://doi.org/10.1109/JSAC.2019.2906789}{\detokenize{10.1109/JSAC.2019.2906789}}}.

\bibitem[Zhuang \em{et~al.}(2020)Zhuang, Ye, Lyu, Cheng, and Ren]{8907851}
Zhuang, W.; Ye, Q.; Lyu, F.; Cheng, N.; Ren, J.
\newblock SDN/NFV-Empowered Future IoV With Enhanced Communication, Computing,
  and Caching.
\newblock {\em Proceedings of the IEEE} {\bf 2020}, {\em 108},~274--291.
\newblock
  doi:{\changeurlcolor{black}\href{https://doi.org/10.1109/JPROC.2019.2951169}{\detokenize{10.1109/JPROC.2019.2951169}}}.

\bibitem[Sun \em{et~al.}(2024)Sun, Cheng, Li, Chen, and Chen]{10387423}
Sun, R.; Cheng, N.; Li, C.; Chen, F.; Chen, W.
\newblock Knowledge-Driven Deep Learning Paradigms for Wireless Network
  Optimization in 6G.
\newblock {\em IEEE Network} {\bf 2024}, {\em 38},~70--78.
\newblock
  doi:{\changeurlcolor{black}\href{https://doi.org/10.1109/MNET.2024.3352257}{\detokenize{10.1109/MNET.2024.3352257}}}.

\bibitem[Wu \em{et~al.}(2024)Wu, Lyu, Ren, Yang, Qian, Gao, and
  Zhang]{10032660}
Wu, F.; Lyu, F.; Ren, J.; Yang, P.; Qian, K.; Gao, S.; Zhang, Y.
\newblock Characterizing Internet Card User Portraits for Efficient Churn
  Prediction Model Design.
\newblock {\em IEEE Transactions on Mobile Computing} {\bf 2024}, {\em
  23},~1735--1752.
\newblock
  doi:{\changeurlcolor{black}\href{https://doi.org/10.1109/TMC.2023.3241206}{\detokenize{10.1109/TMC.2023.3241206}}}.

\bibitem[Wu \em{et~al.}(2023)Wu, Lyu, Wu, Ren, Zhang, and Shen]{9877926}
Wu, F.; Lyu, F.; Wu, H.; Ren, J.; Zhang, Y.; Shen, X.
\newblock Characterizing User Association Patterns for Optimizing Small-Cell
  Edge System Performance.
\newblock {\em IEEE Network} {\bf 2023}, {\em 37},~210--217.
\newblock
  doi:{\changeurlcolor{black}\href{https://doi.org/10.1109/MNET.121.2200089}{\detokenize{10.1109/MNET.121.2200089}}}.

\bibitem[Wang \em{et~al.}(2022)Wang, Han, Li, He, Kumar~Sharma, and
  Chen]{9439054}
Wang, J.; Han, H.; Li, H.; He, S.; Kumar~Sharma, P.; Chen, L.
\newblock Multiple Strategies Differential Privacy on Sparse Tensor
  Factorization for Network Traffic Analysis in 5G.
\newblock {\em IEEE Transactions on Industrial Informatics} {\bf 2022}, {\em
  18},~1939--1948.
\newblock
  doi:{\changeurlcolor{black}\href{https://doi.org/10.1109/TII.2021.3082576}{\detokenize{10.1109/TII.2021.3082576}}}.

\bibitem[Wang \em{et~al.}(2010)Wang, Wu, and Fan]{5586664}
Wang, Q.; Wu, D.O.; Fan, P.
\newblock Delay-Constrained Optimal Link Scheduling in Wireless Sensor
  Networks.
\newblock {\em IEEE Transactions on Vehicular Technology} {\bf 2010}, {\em
  59},~4564--4577.
\newblock
  doi:{\changeurlcolor{black}\href{https://doi.org/10.1109/TVT.2010.2080695}{\detokenize{10.1109/TVT.2010.2080695}}}.

\bibitem[Deng \em{et~al.}(2023)Deng, Zhang, Zhang, Di, Zhang, Poor, and
  Song]{10163760}
Deng, R.; Zhang, Y.; Zhang, H.; Di, B.; Zhang, H.; Poor, H.V.; Song, L.
\newblock Reconfigurable Holographic Surfaces for Ultra-Massive MIMO in 6G:
  Practical Design, Optimization and Implementation.
\newblock {\em IEEE Journal on Selected Areas in Communications} {\bf 2023},
  {\em 41},~2367--2379.
\newblock
  doi:{\changeurlcolor{black}\href{https://doi.org/10.1109/JSAC.2023.3288248}{\detokenize{10.1109/JSAC.2023.3288248}}}.

\bibitem[Yue \em{et~al.}(2024)Yue, Zeng, Liu, Eldar, and Di]{10620366}
Yue, S.; Zeng, S.; Liu, L.; Eldar, Y.C.; Di, B.
\newblock Hybrid Near-Far Field Channel Estimation for Holographic MIMO
  Communications.
\newblock {\em IEEE Transactions on Wireless Communications} {\bf 2024}, pp.
  1--1.
\newblock
  doi:{\changeurlcolor{black}\href{https://doi.org/10.1109/TWC.2024.3433491}{\detokenize{10.1109/TWC.2024.3433491}}}.

\bibitem[Kaul \em{et~al.}(2011)Kaul, Gruteser, Rai, and
  Kenney]{kaul2011minimizing}
Kaul, S.; Gruteser, M.; Rai, V.; Kenney, J.
\newblock Minimizing age of information in vehicular networks.
\newblock  2011 8th Annual IEEE communications society conference on sensor,
  mesh and ad hoc communications and networks. IEEE,  2011, pp. 350--358.

\bibitem[Qi \em{et~al.}(2024{\natexlab{a}})Qi, Wu, Fan, Cheng, Chen, Wang, and
  Letaief]{10663259}
Qi, K.; Wu, Q.; Fan, P.; Cheng, N.; Chen, W.; Wang, J.; Letaief, K.B.
\newblock Deep-Reinforcement-Learning-Based AoI-Aware Resource Allocation for
  RIS-Aided IoV Networks.
\newblock {\em IEEE Transactions on Vehicular Technology} {\bf 2024}, pp.
  1--14.
\newblock
  doi:{\changeurlcolor{black}\href{https://doi.org/10.1109/TVT.2024.3452790}{\detokenize{10.1109/TVT.2024.3452790}}}.

\bibitem[Qi \em{et~al.}(2024{\natexlab{b}})Qi, Wu, Fan, Cheng, Fan, and
  Wang]{10654286}
Qi, K.; Wu, Q.; Fan, P.; Cheng, N.; Fan, Q.; Wang, J.
\newblock Reconfigurable Intelligent Surface Assisted VEC Based on Multi-Agent
  Reinforcement Learning.
\newblock {\em IEEE Communications Letters} {\bf 2024}, {\em 28},~2427--2431.
\newblock
  doi:{\changeurlcolor{black}\href{https://doi.org/10.1109/LCOMM.2024.3451182}{\detokenize{10.1109/LCOMM.2024.3451182}}}.

\bibitem[Shao \em{et~al.}(2024)Shao, Wu, Fan, Cheng, Fan, and Wang]{10636300}
Shao, Z.; Wu, Q.; Fan, P.; Cheng, N.; Fan, Q.; Wang, J.
\newblock Semantic-Aware Resource Allocation Based on Deep Reinforcement
  Learning for 5G-V2X HetNets.
\newblock {\em IEEE Communications Letters} {\bf 2024}, {\em 28},~2452--2456.
\newblock
  doi:{\changeurlcolor{black}\href{https://doi.org/10.1109/LCOMM.2024.3443603}{\detokenize{10.1109/LCOMM.2024.3443603}}}.

\bibitem[Cui \em{et~al.}(2024)Cui, Xiao, Qiong, Pingyi, Qiang, Huiling, and
  Jiangzhou]{10670125}
Cui, Z.; Xiao, X.; Qiong, W.; Pingyi, F.; Qiang, F.; Huiling, Z.; Jiangzhou, W.
\newblock Anti-Byzantine attacks enabled vehicle selection for asynchronous
  federated learning in vehicular edge computing.
\newblock {\em China Communications} {\bf 2024}, {\em 21},~1--17.
\newblock
  doi:{\changeurlcolor{black}\href{https://doi.org/10.23919/JCC.fa.2023-0718.202408}{\detokenize{10.23919/JCC.fa.2023-0718.202408}}}.

\bibitem[Wu \em{et~al.}(2024)Wu, Wang, Ge, Fan, Fan, and Letaief]{10278180}
Wu, Q.; Wang, S.; Ge, H.; Fan, P.; Fan, Q.; Letaief, K.B.
\newblock Delay-Sensitive Task Offloading in Vehicular Fog Computing-Assisted
  Platoons.
\newblock {\em IEEE Transactions on Network and Service Management} {\bf 2024},
  {\em 21},~2012--2026.
\newblock
  doi:{\changeurlcolor{black}\href{https://doi.org/10.1109/TNSM.2023.3322881}{\detokenize{10.1109/TNSM.2023.3322881}}}.

\bibitem[Cao \em{et~al.}(2024)Cao, Gu, Wu, and Wang]{CAO2024103438}
Cao, D.; Gu, N.; Wu, M.; Wang, J.
\newblock Cost-effective task partial offloading and resource allocation for
  multi-vehicle and multi-MEC on B5G/6G edge networks.
\newblock {\em Ad Hoc Networks} {\bf 2024}, {\em 156},~103438.
\newblock
  doi:{\changeurlcolor{black}\href{https://doi.org/https://doi.org/10.1016/j.adhoc.2024.103438}{\detokenize{https://doi.org/10.1016/j.adhoc.2024.103438}}}.

\bibitem[Gu \em{et~al.}(2024)Gu, Wu, Fan, Fan, Cheng, Chen, and
  Letaief]{10745538}
Gu, X.; Wu, Q.; Fan, P.; Fan, Q.; Cheng, N.; Chen, W.; Letaief, K.B.
\newblock DRL-Based Resource Allocation for Motion Blur Resistant Federated
  Self-Supervised Learning in IoV.
\newblock {\em IEEE Internet of Things Journal} {\bf 2024}, pp. 1--1.
\newblock
  doi:{\changeurlcolor{black}\href{https://doi.org/10.1109/JIOT.2024.3492326}{\detokenize{10.1109/JIOT.2024.3492326}}}.

\bibitem[Ji \em{et~al.}(2024)Ji, Wu, Fan, Cheng, Chen, Wang, and
  Letaief]{10697115}
Ji, M.; Wu, Q.; Fan, P.; Cheng, N.; Chen, W.; Wang, J.; Letaief, K.B.
\newblock Graph Neural Networks and Deep Reinforcement Learning Based Resource
  Allocation for V2X Communications.
\newblock {\em IEEE Internet of Things Journal} {\bf 2024}, pp. 1--1.
\newblock
  doi:{\changeurlcolor{black}\href{https://doi.org/10.1109/JIOT.2024.3469547}{\detokenize{10.1109/JIOT.2024.3469547}}}.

\bibitem[Qi \em{et~al.}(2024)Qi, Wu, Fan, Cheng, Chen, and Letaief]{10699378}
Qi, K.; Wu, Q.; Fan, P.; Cheng, N.; Chen, W.; Letaief, K.B.
\newblock Reconfigurable Intelligent Surface Aided Vehicular Edge Computing:
  Joint Phase-Shift Optimization and Multi-User Power Allocation.
\newblock {\em IEEE Internet of Things Journal} {\bf 2024}, pp. 1--1.
\newblock
  doi:{\changeurlcolor{black}\href{https://doi.org/10.1109/JIOT.2024.3470129}{\detokenize{10.1109/JIOT.2024.3470129}}}.

\bibitem[Moltafet \em{et~al.}(2020)Moltafet, Leinonen, and Codreanu]{9217386}
Moltafet, M.; Leinonen, M.; Codreanu, M.
\newblock An Exact Expression for the Average AoI in a Multi-Source M/M/1
  Queueing Model.
\newblock  2020 IEEE 31st Annual International Symposium on Personal, Indoor
  and Mobile Radio Communications,  2020, pp. 1--6.
\newblock
  doi:{\changeurlcolor{black}\href{https://doi.org/10.1109/PIMRC48278.2020.9217386}{\detokenize{10.1109/PIMRC48278.2020.9217386}}}.

\bibitem[Zhang \em{et~al.}(2023)Zhang, Chen, Chen, Tian, Jia, Wang, and
  Wu]{10198349}
Zhang, T.; Chen, S.; Chen, Z.; Tian, Z.; Jia, Y.; Wang, M.; Wu, D.O.
\newblock AoI and PAoI in the IoT-Based Multisource Status Update System:
  Violation Probabilities and Optimal Arrival Rate Allocation.
\newblock {\em IEEE Internet of Things Journal} {\bf 2023}, {\em
  10},~20617--20632.
\newblock
  doi:{\changeurlcolor{black}\href{https://doi.org/10.1109/JIOT.2023.3297617}{\detokenize{10.1109/JIOT.2023.3297617}}}.

\bibitem[Inoue \em{et~al.}(2019)Inoue, Masuyama, Takine, and Tanaka]{8820073}
Inoue, Y.; Masuyama, H.; Takine, T.; Tanaka, T.
\newblock A General Formula for the Stationary Distribution of the Age of
  Information and Its Application to Single-Server Queues.
\newblock {\em IEEE Transactions on Information Theory} {\bf 2019}, {\em
  65},~8305--8324.
\newblock
  doi:{\changeurlcolor{black}\href{https://doi.org/10.1109/TIT.2019.2938171}{\detokenize{10.1109/TIT.2019.2938171}}}.

\bibitem[Zou \em{et~al.}(2021)Zou, Ozel, and Subramaniam]{9347556}
Zou, P.; Ozel, O.; Subramaniam, S.
\newblock Optimizing Information Freshness Through Computation–Transmission
  Tradeoff and Queue Management in Edge Computing.
\newblock {\em IEEE/ACM Transactions on Networking} {\bf 2021}, {\em
  29},~949--963.
\newblock
  doi:{\changeurlcolor{black}\href{https://doi.org/10.1109/TNET.2021.3053937}{\detokenize{10.1109/TNET.2021.3053937}}}.

\bibitem[Chu \em{et~al.}(2023)Chu, Wu, Fan, and Li]{10419638}
Chu, J.; Wu, Q.; Fan, Q.; Li, Z.
\newblock Enhanced C-V2X Mode 4 to Optimize Age of Information and Reliability
  for IoV.
\newblock  2023 IEEE 23rd International Conference on Communication Technology
  (ICCT),  2023, pp. 1082--1086.
\newblock
  doi:{\changeurlcolor{black}\href{https://doi.org/10.1109/ICCT59356.2023.10419638}{\detokenize{10.1109/ICCT59356.2023.10419638}}}.

\bibitem[Peng \em{et~al.}(2021)Peng, Jiang, Zhang, and Xu]{peng2020age}
Peng, F.; Jiang, Z.; Zhang, S.; Xu, S.
\newblock Age of Information Optimized MAC in V2X Sidelink via Piggyback-Based
  Collaboration.
\newblock {\em IEEE Transactions on Wireless Communications} {\bf 2021}, {\em
  20},~607--622.
\newblock
  doi:{\changeurlcolor{black}\href{https://doi.org/10.1109/TWC.2020.3027353}{\detokenize{10.1109/TWC.2020.3027353}}}.

\bibitem[Hu \em{et~al.}(2022)Hu, Xiong, Lu, Gao, Fan, and Letaief]{9442821}
Hu, H.; Xiong, K.; Lu, Y.; Gao, B.; Fan, P.; Letaief, K.B.
\newblock $\alpha$-$\beta$ AoI Penalty in Wireless-Powered Status Update
  Networks.
\newblock {\em IEEE Internet of Things Journal} {\bf 2022}, {\em 9},~474--484.
\newblock
  doi:{\changeurlcolor{black}\href{https://doi.org/10.1109/JIOT.2021.3084161}{\detokenize{10.1109/JIOT.2021.3084161}}}.

\bibitem[Lin and Liao(2024)]{10683037}
Lin, C.Y.; Liao, W.
\newblock Energy-aware Age of Information (AoI) Minimization for Internet of
  Things in NOMA-based LEO Satellite Networks.
\newblock  2024 IEEE 99th Vehicular Technology Conference (VTC2024-Spring),
  2024, pp. 1--5.
\newblock
  doi:{\changeurlcolor{black}\href{https://doi.org/10.1109/VTC2024-Spring62846.2024.10683037}{\detokenize{10.1109/VTC2024-Spring62846.2024.10683037}}}.

\bibitem[Prabhakaran \em{et~al.}(2023)Prabhakaran, Chavhan, Kumar, and
  Rodrigues]{10193386}
Prabhakaran, P.; Chavhan, S.; Kumar, M.; Rodrigues, J.J.P.C.
\newblock ML-based Minimization of AoI in a Vehicular Communication Network.
\newblock  2023 8th International Conference on Smart and Sustainable
  Technologies (SpliTech),  2023, pp. 1--6.
\newblock
  doi:{\changeurlcolor{black}\href{https://doi.org/10.23919/SpliTech58164.2023.10193386}{\detokenize{10.23919/SpliTech58164.2023.10193386}}}.

\bibitem[Parvini \em{et~al.}(2023)Parvini, Javan, Mokari, Abbasi, and
  Jorswieck]{parvini2023aoi}
Parvini, M.; Javan, M.R.; Mokari, N.; Abbasi, B.; Jorswieck, E.A.
\newblock AoI-Aware Resource Allocation for Platoon-Based C-V2X Networks via
  Multi-Agent Multi-Task Reinforcement Learning.
\newblock {\em IEEE Transactions on Vehicular Technology} {\bf 2023}, {\em
  72},~9880--9896.
\newblock
  doi:{\changeurlcolor{black}\href{https://doi.org/10.1109/TVT.2023.3259688}{\detokenize{10.1109/TVT.2023.3259688}}}.

\bibitem[Emara \em{et~al.}(2020)Emara, Filippou, and Sabella]{emara2020mec}
Emara, M.; Filippou, M.C.; Sabella, D.
\newblock MEC-Enhanced Information Freshness for Safety-Critical C-V2X
  Communications.
\newblock  2020 IEEE International Conference on Communications Workshops (ICC
  Workshops),  2020, pp. 1--5.
\newblock
  doi:{\changeurlcolor{black}\href{https://doi.org/10.1109/ICCWorkshops49005.2020.9145387}{\detokenize{10.1109/ICCWorkshops49005.2020.9145387}}}.

\bibitem[Mlika and Cherkaoui(2022)]{mlika2022deep}
Mlika, Z.; Cherkaoui, S.
\newblock Deep deterministic policy gradient to minimize the age of information
  in cellular V2X communications.
\newblock {\em IEEE Transactions on Intelligent Transportation Systems} {\bf
  2022}, {\em 23},~23597--23612.

\bibitem[Pokhrel and Mandjes(2019)]{TMC_Pokhrel}
Pokhrel, S.R.; Mandjes, M.
\newblock Improving Multipath TCP Performance over WiFi and Cellular Networks:
  An Analytical Approach.
\newblock {\em IEEE Transactions on Mobile Computing} {\bf 2019}, {\em
  18},~2562--2576.
\newblock
  doi:{\changeurlcolor{black}\href{https://doi.org/10.1109/TMC.2018.2876366}{\detokenize{10.1109/TMC.2018.2876366}}}.

\bibitem[Pokhrel and Choi(2020)]{TCOM_Pokhrel}
Pokhrel, S.R.; Choi, J.
\newblock Federated Learning With Blockchain for Autonomous Vehicles: Analysis
  and Design Challenges.
\newblock {\em IEEE Transactions on Communications} {\bf 2020}, {\em
  68},~4734--4746.
\newblock
  doi:{\changeurlcolor{black}\href{https://doi.org/10.1109/TCOMM.2020.2990686}{\detokenize{10.1109/TCOMM.2020.2990686}}}.

\end{thebibliography}

\end{document}